\documentclass[conference]{IEEEtran}
\usepackage{graphicx}
\usepackage{amsmath,amssymb,dsfont,bbm,epstopdf,pgfplots,mathtools,enumitem,mathrsfs, bbm, algpseudocode , graphicx, dblfloatfix, booktabs, physics, algorithm,algcompatible, derivative, physics} 
\usepackage[normalem]{ulem}
\usepackage{color}
\usepackage{cite}
\usepackage{graphics}
\usepackage{tikz}
\usepackage{pgfplots}
\usepackage{subcaption}
\usepackage[utf8]{inputenc} 
\usepackage[T1]{fontenc}
\usepackage[left=0.6in,right=0.6in,top=0.71in, bottom = 0.94in]{geometry}
\usetikzlibrary{shapes, arrows, decorations.markings, arrows.meta}
\usetikzlibrary{patterns} 
\allowdisplaybreaks
\graphicspath{{.}{./Figures/}} 
\allowdisplaybreaks[4] 

\newtheorem{lemma}{Lemma}

\newtheorem{theorem}{Theorem}
\newtheorem{proposition}{Proposition}

\newtheorem{definition}{Definition}

\newcommand{\R}{{\mathbb R}}

\renewcommand{\Pr}{{\mathbb{P}}}

\newcommand{\vect}[1]{\boldsymbol{#1}}
\newcommand{\vectsf}[1]{\boldsymbol {\mathsf {#1}}}

\definecolor{ForestGreen}{rgb}{0.0, 0.5, 0.0}

\renewcommand{\P}{\mathsf{P}}
\newcommand{\D}{\mathsf{D}}
\newcommand{\C}{\mathsf{C}}
\renewcommand{\R}{\mathsf{R}}
\renewcommand{\P}{\mathsf{P}}

\newcommand{\U}{\mathsf{U}}
\newcommand{\e}{\mathsf{e}}
\newcommand{\s}{\mathsf{s}}

\newcommand{\pli}{p_{\ell_i}}

\newcommand{\Bdet}{\mathcal B_{\text{detect}}}
\newcommand{\Barr}{\mathcal B_{\text{arrival}}}
\newcommand{\Bdec}{\mathcal B_{\text{decode}}}

\newcommand{\Bdt}{B_{\text{dt}}}
\newcommand{\vbj}{\vect V_{b,j}}
\newcommand{\tepf}{\tilde \epsilon_{\U,1}}
\newcommand{\teps}{\tilde \epsilon_{\U,2}}

\newcommand{\I}{\vectsf{I}}
\newcommand{\bc}{\beta_u}
\newcommand{\bsf}{\beta_{s,1}}
\newcommand{\bss}{\beta_{s,2}}
\newcommand{\alc}{\alpha_u}
\newcommand{\alsf}{\alpha_{s,1}}
\newcommand{\alss}{\alpha_{s,2}}
\newcommand{\sigbj}{\sigma^2_{b,j}}
\begin{document}
\title{ A MIMO ISAC System for Ultra-Reliable and Low-Latency Communications}
\author{\IEEEauthorblockN{Homa Nikbakht$^{1,2}$, Yonina C.~Eldar$^2$, and H.~Vincent Poor$^1$}
	\IEEEauthorblockA{$^1$Princeton University,   $\quad ^2$Weizmann Institute of Science \\
		\{homa, poor\}@princeton.edu, yonina.eldar@weizmann.ac.il}}
\maketitle
\begin{abstract}
In this paper, we propose a bi-static multiple-input multiple-output (MIMO) integrated sensing and communication (ISAC) system to detect the arrival of ultra-reliable and low-latency communication (URLLC) messages and prioritize their delivery. In this system, a dual-function base station (BS) communicates with a user equipment (UE) and a sensing receiver (SR) is deployed to collect echo signals reflected from a target of interest.  The BS regularly transmits messages of enhanced mobile broadband (eMBB) services to the UE. During each eMBB transmission, if the SR senses the presence of a target of interest, it immediately triggers the transmission of an additional URLLC message. To reinforce URLLC transmissions, we propose a dirty-paper coding (DPC)-based technique that mitigates the interference of both eMBB and sensing signals. For this system, we formulate the rate-reliability-detection trade-off in the finite blocklength regime by evaluating  the communication rate of  the eMBB transmissions, the reliability of  the URLLC transmissions and the  probability of  the target detection. Our numerical analysis show that  our proposed DPC-based ISAC scheme significantly outperforms power-sharing based ISAC  and traditional time-sharing schemes. In particular, it achieves higher eMBB transmission rate while satisfying both URLLC and sensing constraints.
\end{abstract}
\section{Introduction}

Integrated sensing and communication (ISAC) is enabled by higher frequency bands, wider bandwidths and denser distributions of massive antenna arrays.  Such integration is mutually beneficial to sensing and communication tasks thus offering various use cases for autonomous driving, smart factories, and other environmental-aware scenarios in 6G  \cite{Visa2024, Luo2024, Liu2022, Li2024, YLiu2024}.  Many of these use cases also require ultra-reliable and low-latency communication (URLLC), a feature introduced in 5G and expected to be further advanced in 6G \cite{Mahmood2023, Behdad2024, Qin2025, Zhao2022}. 

Despite great progress to achieve the required latency and reliability, 5G URLLC still does not meet all key performance metrics needed for diverse mission-critical applications. One of the main challenges arises from the random generation nature of URLLC services, as their generation is often linked to the occurrence of critical, time-sensitive events in the environment. 
Another challenge is the coexistence of URLLC services with other 5G/6G services such as enhanced mobile broadband (eMBB) type services that depend largely on high transmission rate and are less sensitive to delay. Different coexistence strategies have been studied in the literature \cite{HomaEntropy2022, Song2019,Interdonato2023, HomaGlobecom2023, Wang2024}. For example, the work in \cite{Song2019} proposes a puncturing strategy in which  the on-going eMBB transmission stops upon the arrival of URLLC messages. The work in \cite{Interdonato2023} shows that a superposition coding strategy in which the transmitter simply sends a linear combination of eMBB and URLLC signals outperforms the puncturing strategy. A  dirty-paper coding (DPC) \cite{Costa1983, HomaISIT2022} based joint transmission strategy is proposed in \cite{HomaGlobecom2023} which also outperforms the puncturing technique.  Theses studies either assume a deterministic model or a random model with Bernoulli distribution for the arrival of URLLC messages, which have shortcomings in offering a practical model for the stochastic nature of this type of services. 

In this work, we propose a bi-static MIMO ISAC-enabled URLLC system that supports the joint transmission of eMBB and URLLC services. This system uses  its own sensory data to trigger URLLC transmissions with no assumption on their arrival  distribution. The setup consists of one base station (BS), one user equipment (UE), one target and one sensing receiver (SR).
 The BS accommodates the transmission of both  eMBB and URLLC type services. 
 The eMBB message arrives at the beginning of the transmission slot and its transmission lasts over the entire slot. We divide the eMBB slot into smaller blocks. In each block, the BS transmits dual-function signals enabling simultaneous communication and sensing tasks.  In each block,  if the SR senses the presence of the target, it triggers the transmission of an additional  URLLC message over the next immediate block. In blocks with no URLLC transmission, the dual function signal is generated using DPC to precancel the interference of sensing signal from the eMBB transmission \cite{Li2023Globecom}. In blocks with joint URLLC and eMBB transmissions,  to increase the reliability of the URLLC transmission, we also propose a  DPC based method to generate the dual function signal. In this method, we first precancel the interference of  the sensing signal from the eMBB transmission, and then precancel the interference of both eMBB and sensing signals from the URLLC transmission. After each block, the UE attempts to decode a URLLC message, and after the entire transmission slot it decodes the eMBB message. 
 
 For this system, we formulate the rate-reliability-detection trade-off in the finite blocklength regime \cite{Nikbakht2024}. Specifically,  we upper bound the eMBB rate while treating the interference of URLLC as  noise. We measure the reliability of the almost interference-free URLLC transmissions by  evaluating the corresponding decoding error probability. Finally, we calculate the target detection probability under the interference of both URLLC and eMBB. 
Through numerical analyses we show that  our proposed DPC-based ISAC scheme outperforms the power-sharing and the time-sharing schemes by achieving higher eMBB rate while accommodating the URLLC constraints and the sensing constraints of the target detection. 
\section{Problem Setup}
\begin{figure}[t]
\center
\includegraphics[width=0.44\textwidth]{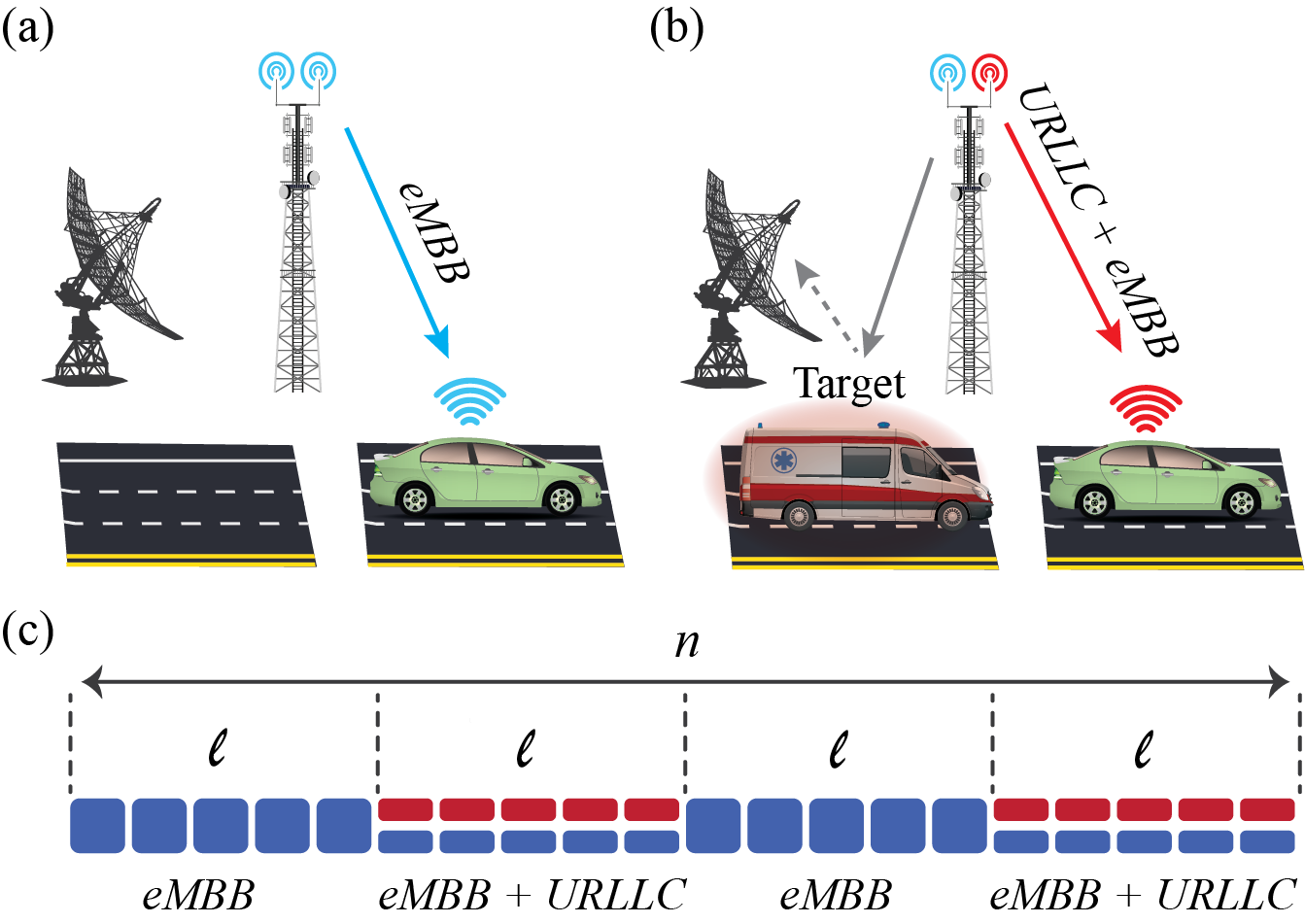}
\vspace{-0.1cm}
\caption{An illustration of the system model: (a) no target is detected, (b) detection of a target, (c) example of transmission blocks with $\eta = 4$.} 
\label{fig1}
\vspace{-0.4cm}
\end{figure}
Consider a  bi-static MIMO ISAC system where a BS is communicating with  a UE and simultaneously wishes to sense a target of interest. A SR is deployed to collect echo signals reflected from the target\footnote{We assume that the UE is not within the  SR sensing range. Hence, the SR does not receive echo signals from the UE. We also assume that the SR and the BS can communicate over an interference-free backhaul link.}. The BS is equipped with $t$ transmit antennas, the UE and the SR each are equipped with $r$ receive antennas. The BS communicates both eMBB and URLLC type messages to the UE. Assume $n$ is the total communication frame length. The eMBB message $m_{\e}$ is uniformly distributed over a given set $\mathcal M_{\e} : = \{1, \ldots, M_{\e}\}$ and is sent over the entire transmission interval $n$.
We divide $n$ into $\eta$ blocks each of length $\ell$ (i.e., $n = \eta \cdot \ell$), as in Fig.~\ref{fig1}. Denote by $P_{b,\D}$ as the target detection probability in block~$b$.
 In each block~$b \in [\eta]$, with $[\eta]:=\{1,\ldots, \eta\}$, the BS transmits an additional URLLC message $m_{b,\U}$  to the UE with probability $P_{b-1,\D}$ which is the target detection probability in the previous block~$b-1$. With probability $1-P_{b-1, \D}$ no URLLC message is generated. The message $m_{b,\U}$ is uniformly distributed over a given set $\mathcal M_{\U} : = \{1, \ldots, M_{\U}\}$. For each $b \in [\eta]$, if a URLLC message is generated, then we set $A_b =1$, and otherwise we set $A_b = 0$. Note that $A_1 = 0$ with probability $1$. 
Define
\begin{IEEEeqnarray}{rCl}\label{eq:barr}
\mathcal B_{\text{arrival}} : = \{ b \in [\eta]: A_b = 1\},
\end{IEEEeqnarray}
as the set of blocks in which an additional URLLC message is generated. 

\subsection{Communication Received Signal at the UE}
At the end of each block~$b \in [\eta]$, the UE receives the signal $\vectsf Y_{b,c} = [\vect Y_{b,c,1} \ldots \vect Y_{b,c,r} ] \in \mathbb C^{\ell \times r}$ from the BS. Assume a MIMO memoryless Gaussian quasi-static fading channel. The channel input-output relation in each block~$b \in [\eta]$ is given by
\begin{equation} \label{eq:channel}
\vectsf Y_{b,c} =(\vectsf X_{b,c} + \vectsf X_{b,s})  \vectsf H_{b,c} \ + \vectsf N_{b,c},
\end{equation}
where $\vectsf X_{b,c} \in \mathbb C^{\ell \times t}$ is the communication signal,  $\vectsf X_{b,s} \in \mathbb C^{\ell \times t}$ is the sensing signal,  $\vectsf Y_{b,c} \in \mathbb C^{\ell\times r}$ is the communication channel output, $\vectsf H_{b,c} \in \mathbb C^{t \times r}$ is the communication channel matrix, and $ \vectsf N_{b,c} \in \mathbb C^{\ell \times r}$ is additive noise at the UE whose entries are i.i.d. $\mathcal {N} (0,1)$ and is independent of $\vectsf H_{b,c}$. 
Let $\vectsf X_b = \vectsf X_{b,c} + \vectsf X_{b,s}$, and denote  $\vectsf X : = [\vectsf X_1, \ldots, \vectsf X_{\eta}]$. The input matrix $\vectsf X$ is admissible if it belongs to the following set:
\begin{equation}
\mathcal P_X (\P) := \{ \vectsf X \in \mathbb C^{n \times t} : \text{Tr} \left(\vectsf X \vectsf X^H\right) \le n \P \}, \label{eq:1}
\end{equation}
that implies a power constraint on the input matrix $\vectsf X$ by upper bounding the trace of $\vectsf X \vectsf X^H$ with $n \P$. 
Assume that the channel state information is known at both the BS and the UE. We next introduce a communication channel code notation for this setting.  
\begin{definition}\label{def:code}
A communication channel code $(n, M_{\e}, M_{\U}, \epsilon_{\e}^{(n)}, \{\epsilon_{b,\U}\}_{b = 1}^{\eta})$ consists of:
\begin{itemize}
\item  An encoder $f_{\e}: \mathcal M_{\e} \times \mathbb C^{t\times r} \to \mathbb C^{t\times r}$ that at each block~$b \notin \mathcal B_{\text{arrival}}$ maps the message $m_{\e} \in \mathcal M_{\e}$ and the channel $\vectsf H_{b,c}$ to a codeword $\vectsf X_{b,c} = f_{\e}(m_{\e}, \vectsf H_{b,c})$ 
for each $m_{\e} \in \mathcal M_{\e}$ and $\vectsf H_{b,c} \in \mathbb C^{t\times r}$. 

\item An encoder $f_{b}: \mathcal M_{\e} \times \mathcal M_{\U} \times \mathbb C^{t\times r} \to \mathbb C^{t\times r}$ that at each block~$b \in \mathcal B_{\text{arrival}}$ maps the message pair $(m_{\e}, m_{b,\U})$  and the channel $\vectsf H_{b,c}$ to a codeword $\vectsf X_{b,c} = f_b(m_{\e}, m_{b,\U}, \vectsf H_{b,c})$
for each $m_{\e} \in \mathcal M_{\e}$, $m_{b,\U} \in \mathcal M_{\U}$ and $\vectsf H_{b,c} \in \mathbb C^{t\times r}$.

\item A decoder $g_{b,u}: \mathbb C^{\ell \times r} \times \mathbb C^{t\times r} \to \mathcal M_{\U}$ that in each block~$b \in \mathcal B_{\text{arrival}}$ produces 
\begin{equation}
\hat m_{b, \U} = g_{b,u} (\vectsf Y_{b,c}, \vectsf H_{b,c}),
\end{equation}
and produces $\hat m_{b, \U}  = 0$ if $b \notin \mathcal B_{\text{arrival}}$. For each message $m_{b,\U}$, the average error probability is defined as
\begin{IEEEeqnarray}{rCl}
\epsilon_{b,\U}&: =& P_{b-1,\D} \Pr [\hat m_{b, \U} \neq m_{b, \U} | b \in \Barr]\notag \\
&& + (1-P_{b-1,\D}) \Pr [\hat m_{b, \U} \neq 0 | b \notin \Barr].
\end{IEEEeqnarray}

\item A decoder $g_{\e}:  \mathbb C^{n \times r} \times \mathbb C^{t\times r} \to \mathcal M_{\e}$ that at the end of the entire $n$ channel uses produces
\begin{equation}
\hat m_{\e} = g_{\e} (\vectsf Y_{1,c}, \ldots, \vectsf Y_{\eta,c}, \vectsf H_{1,c}, \ldots, \vectsf H_{\eta,c}).
\end{equation}
 The average error probability of the eMBB message $m_e$ is defined as 
 \begin{equation}
 \epsilon_{\e}^{(n)} : = \Pr [ \hat m_{\e} \neq m_{\e}].
\end{equation}
\end{itemize}
\end{definition}

\subsection{Target Echo Signal Model}
In block~$b \in [\eta]$, after transmitting $\vectsf X_b$, the SR receives the reflected echo matrix $\vectsf Y_{b,s} = [\vect Y_{b,s,1} \ldots \vect Y_{b,s,r} ] \in \mathbb C^{\ell \times r}$. 
The target detection problem in each block is formulated by defining the following two hypotheses: 
\begin{IEEEeqnarray}{rCl}
\mathcal H_0&:& \vectsf Y_{b,s} = \vectsf N_{b,s},  \\
\mathcal H_1&:& \vectsf Y_{b,s} = (\vectsf X_{b,c} + \vectsf X_{b,s})  \vectsf H_{b,s}  + \vectsf N_{b,s}, 
\end{IEEEeqnarray}
where $\vectsf H_{b,s} \in \mathbb C^{t \times r} $ is the target channel response matrix and $\vectsf N_{b,s} \in \mathbb C^{\ell \times r}$ is additive noise matrix with each entry having zero mean and unit variance. The optimal detector for this problem is the likelihood ratio test (LRT) \cite{Poor} and is given by 
\begin{equation}
T_b = \log  \left ( \prod_{j = 1}^{q} \frac{f\left (\vect Y_{b,s,j}  | \mathcal H_1\right)}{f\left(\vect Y_{b,s,j} | \mathcal H_0\right)} \right) \underset{{\mathcal H_0}}{\overset{\mathcal H_{1}}{\gtrless}} \delta
\end{equation}
where   $f\left (\vect Y_{b,s,j} | \mathcal H_0\right)$ and $f\left (\vect Y_{b,s,j} | \mathcal H_1\right)$ are the probability density functions (pdf) of the observation vector under the null and alternative hypotheses, respectively, and $q:=\min \{r,t\}$. Denote by $P_{b,\D}$ and $P_{b,\text{FA}}$  the target detection and false alarm probabilities of block~$b$, respectively. These probabilities are given by
\begin{subequations}
\begin{IEEEeqnarray}{rCl}
P_{b,\D} &=& \Pr [T_b > \delta | \mathcal H_1], \\
P_{b,\text{FA}} & = & \Pr [T_b > \delta | \mathcal H_0].
\end{IEEEeqnarray}
\end{subequations}

In our analysis, the threshold  $\delta$ is  set to ensure the desired probability of false alarm. In each block~$b$, we assume that the SR has the full knowledge of the sensing channel state information and the sensing signal $\vectsf X_{b,s}$ but the communication signal $\vectsf X_{b,c}$ is unknown to the SR. 

\begin{proposition}\label{prop1}
Given $n, \eta$ and $\P$, let $\R_e := \frac{\log M_{\e}}{n}$ be the eMBB transmission rate.  The rate-reliability-detection trade-off  is
\begin{subequations}
\begin{IEEEeqnarray}{rCl}
\max_{f_{\vectsf X} (\vectsf x) \in \mathcal P_X(\P) } \quad && \R_{\e}\\
\text{subject to}\; \quad && \epsilon_{\e}^{(n)} \le \epsilon_{\e},\\
&& \epsilon_{b, \U} \le \epsilon_{\U}, \quad \forall b \in [\eta], \\
&& P_{b, \D}  \ge \P_{\D}, \quad \forall b \in [\eta],
\end{IEEEeqnarray}
\end{subequations}
where $\mathcal P_X(\P)$ is defined in \eqref{eq:1}. 
\end{proposition}

\section{Coding Scheme}
 In each block $b \in [\eta]$, codewords are generated such that the total transmit power is upper bounded by $\ell \P$ thus satisfying the power constraint \eqref{eq:1} over the $\eta$ blocks. In each block~$b \notin \Barr$,  we employ a DPC technique with parameter $\alsf \in [0,1]$ to precancel the interference of the sensing signal from the eMBB transmission. In each block~$b \in \Barr$ where the BS has both eMBB and URLLC messages to send, we employ a DPC technique to improve the reliability of the URLLC transmission. 
In this technique, we first precancel the interference of  the sensing signal from the eMBB transmission using a DPC with parameter $\alss \in [0,1]$, and then precancel the interference of both eMBB and sensing signals from the URLLC transmission using a DPC technique with parameter $\alc \in [0,1]$.
\subsection{Codebook construction}
Choose $\bc, \bsf, \bss \in [0,1]$.  To improve the finite blocklength performance, all codewords are uniformly distributed on the power shell. Denote a centered $\ell$-dimensional  sphere of radius $r$ by $\mathsf S_{\ell}( r)$. For each block $b \in [\eta]$ and for each $j \in [t]$, generate the following codewords.
\begin{itemize}
\item For each $v \in [ M_v]$ and each  realization  $m \in [ M_{\U}]$, generate  codewords $\vect V_{b,j}(m,v)$    by picking them uniformly over $\mathsf S_{\ell}\left(\sqrt{\ell (\bc + \alc^2(1-\bc)) \P}\right)$. 
\item For each $s \in [M_{\s}]$ and each realization $ m' \in [M_{\e}]$,  randomly draw two codewords: a codeword $\vect S_{b,j}^{(1)}(m',s)$ uniformly distributed  on $\mathsf S_{\ell}\left (\sqrt{\ell (\bsf + \alsf^2(1-\bsf)) \P}\right)$; and a codeword $\vect S_{b,j}^{(2)}(m',s)$ uniformly distributed  on  $\mathsf S_{\ell} \left (\sqrt{\ell (1-\bc)(\bss + \alss^2(1-\bss)) \P}\right)$. 
\item For each $s \in [M_{\s}]$,  randomly draw two codewords: a codeword $\vect X_{b,j}^{(\s,1)}(s)$ uniformly distributed  on $\mathsf S_{\ell} \left (\sqrt{\ell (1-\bsf) \P}\right)$; and a codeword $\vect X_{b,j}^{(\s,2)}(s)$ uniformly distributed  on $\mathsf S_{\ell} \left (\sqrt{\ell(1-\bc)(1-\bss) \P}\right)$. 
\end{itemize}
All codewords are chosen independently of each other. 

\subsection{Encoding}\label{sec:enc1}
\subsubsection{Encoding at each block~$b \notin \Barr$} 
In each block~$b \notin \Barr$ and each $j \in [t]$, the BS first picks its sensing signal $\vect X_{b,j}^{(s,1)}(s)$ and then uses  DPC to encode its eMBB message $m_{\e}$  while precanceling the interference of its own sensing signal. Specifically, it chooses an index $s$ such that the 
\begin{equation} \label{eq:xe1}
\vect X_{b,j}^{(\e,1)} :  = \vect S_{b,j}^{(1)} (m_{\e}, s )- \alsf  \vect X_{b,j}^{(\s,1)}
\end{equation}
 lies in the set $\mathcal D(\ell \bsf \P, \zeta_{s,1})$ for a given $\zeta_{\s,1}> 0$ where 
\begin{IEEEeqnarray}{rCl}\label{eq:di}
\mathcal D(a, \zeta) := \left \{ \vect x: a -\zeta \le \left\| \vect x\right\|^2 \le a \right \}. \IEEEeqnarraynumspace
\end{IEEEeqnarray}
For simplicity, we assume that at least one such a codeword exists. If multiple  such codewords  exist, the index $s^*$ is chosen at random from this set, and the BS sends $\vectsf X_b = [\vect X_{b,1}, \ldots, \vect X_{b, t}]$ with
\begin{equation}
\vect{X}_{b,j}= \vect X_{b,j}^{(\e,1)} + \vect X_{b,j}^{(\s,1)}, \quad j \in [t].
\end{equation}
\subsubsection{Encoding at each block~$b \in \Barr$} 
 In each block~$b \in  \Barr$, the BS has both eMBB and URLLC messages to send. For each $j \in [t]$,  it first picks its sensing signal $\vect X_{b,j}^{(\s,2)}(s)$. It then uses DPC to encode its eMBB message $m_{\e}$ while precanceling the interference of its own sensing signal. More specifically it chooses an index $s$ such that  
\begin{equation} \label{eq:xe1}
\vect X_{b,j}^{(\e,2)} :  = \vect S_{b,j}^{(2)} (m_{\e}, s )- \alss  \vect X_{b,j}^{(\s,2)}
\end{equation}
 lies in the set $\mathcal D \left (\ell \bss(1-\bc) \P, \zeta_{\s,2}\right)$
for a given $\zeta_{\s,2}> 0$. 
  Then it employs DPC to encode $m_{b, \U}$  while precanceling the interference of its own sensing and eMBB signals  $\vect X_{b,j}^{(\e,2)} + \vect X_{b,j}^{(\s,2)}$. 
Specifically, it chooses an index $v$ such that the 
sequence 
\begin{equation} \label{eq:x21}
\vect X_{b,j}^{(\U)} :  = \vect V_{b,j} (m_{b, \U}, v )- \alc  \left (\vect X_{b,j}^{(\e,2)} + \vect X_{b,j}^{(\s,2)}\right)
\end{equation}
 lies in the set $\mathcal D\left (\ell \bc \P, \zeta_u\right )$
for a given $\zeta_u> 0$. 
If multiple  such codewords  exist, indices  $s^*$ and $v^\star$ are chosen at random from these sets, and the BS sends $\vectsf X_b = [\vect X_{b,1}, \ldots, \vect X_{b, t}]$ with
\begin{equation}
\vect{X}_{b,j}= \vect X_{b,j}^{(\U)} + \vect X_{b,j}^{(\e,2)} + \vect X_{b,j}^{(\s,2)}, \quad j \in [t].
\end{equation}




%




\subsection{Decoding} 
  
\subsubsection{Decoding of URLLC Messages} At the end of each block~$b \in [\eta]$, the UE observes the channel outputs $\vectsf Y_{b,c}$:
\begin{IEEEeqnarray}{rCl}
\vectsf Y_{b,c} = \begin{cases} 
( \vectsf X_{b}^{(\e,1)} + \vectsf X_{b}^{(\s,1)}) \vectsf H_{b,c}+ \vectsf N_{b,c} \quad & \text{if}\; b \notin \mathcal B_{\text{arrival}}  \\
(\vectsf X_{b}^{(\U)} + \vectsf X_{b}^{(\e,2)} + \vectsf X_{b}^{(\s,2)})\vectsf H_{b,c}+ \vectsf N_{b,c} \quad & \text{o.w.} 
 \end{cases}\notag  \\
\end{IEEEeqnarray}
Define the information density metric between $\vectsf y_{b,c}$ and $\vectsf v_b: = [\vect v_{b,1}, \ldots, \vect v_{b,q}]$ by:
\begin{equation} \label{eq:ibu}
i^{(\U)}_b  (\vectsf v_b; \vectsf y_{b,c} ) := \log \frac{f_{\vectsf Y_{b,c}| \vectsf V_{b}} (\vectsf y_{b,c}| \vectsf v_{b})}{f_{\vectsf Y_{b,c}}(\vectsf y_{b,c})}. 
\end{equation}

The UE  then chooses  the pair
\begin{equation}
(m^*,v^*) =\text{arg} \max_{ m, v}  i^{(\U)}_b  (\vectsf v_b; \vectsf y_{b,c} ).
\end{equation}
Given a threshold $\delta_{\U}$, if for this pair 
\begin{equation}
 i^{(\U)}_b  (\vectsf v_b; \vectsf y_{b,c} ) > \delta_{\U}
 \end{equation}
 the UE chooses $(\hat m_{b,\U},\hat v)= (m^*,v^*)$. Otherwise the UE declares that no URLLC message has been received and indicates it by setting $\hat m_{b,\U}=0$. 
Define
\begin{IEEEeqnarray}{rCl} 
\mathcal B_{\text{detect}} &:=& \{b \in [\eta]: \hat m_{b,\U} \neq 0\}, \label{eq:bdet}\\
\mathcal B_{\text{decode}} &: =& \{b \in \mathcal B_{\text{detect}}: \hat m_{b,\U} = m_{b,\U}\},\label{eq:bdec}
\end{IEEEeqnarray}
where $\Bdet$ denotes the set of blocks in which a URLLC message is detected and $\Bdec$ denotes the set of blocks in which a URLLC message is decoded correctly. 

\subsubsection{Decoding the eMBB Message} \label{sec:eMBBTIN}
The UE decodes its eMBB message based on the output of the entire $\eta$ blocks. Let $\vectsf Y_c : = [\vectsf Y_{1,c}, \ldots, \vectsf Y_{\eta, c}]$. Given that the URLLC messages interfere on the eMBB transmissions, the UE treats URLLC transmissions as noise. Therefore, the decoding of the eMBB message depends on the detection of URLLC messages sent over the $\eta$ blocks. Let $B_{\text{dt}}$ be the realization of the set $ \mathcal B_{\text{detect}}$ defined in \eqref{eq:bdet}. Also, let $\vectsf s_{\e,1} : =  \{\vectsf s_{b}^{(1)}\}_{b \notin B_{\text{dt}} }$, and $\vectsf s_{\e,2} : =  \{\vectsf s_{b}^{(2)}\}_{b \in  B_{\text{dt}} }$. Given $B_{\text{dt}}$, the UE decodes its eMBB message based on the outputs of the entire $n$ channel uses by  looking for an index pair $(m',s)$ such that its corresponding codewords $\left \{ \vectsf s_{\e,2} (m',s),  \vectsf s_{\e,1} (m',s) \right \}$ maximize 
\begin{IEEEeqnarray}{rCl}
\lefteqn{i^{(\e)} \left ( \vectsf s_{\e,2} ,  \vectsf s_{\e,1} ;  \vectsf y_c| \mathcal B_{\text{detect}} = B_{\text{dt}} \right):=}\notag \\ &&
 \log \hspace{-0.15cm}\prod_{b\notin  B_{\text{dt}}  }\hspace{0cm}  \frac{f_{\vectsf Y_{b,c}| \vectsf S_{b}^{(1)}} (\vectsf y_{b,c}| \vectsf s_{b}^{(1)})}{f_{\vectsf Y_{b,c}}(\vectsf y_{b,c})} +  \log \hspace{-0.15cm}\prod_{b\in  B_{\text{dt}} } \hspace{0cm}  \frac{f_{\vectsf Y_{b,c} | \vectsf S_{b}^{(2)}} (\vectsf y_{b,c}| \vectsf s_{b}^{(1)})}{f_{\vectsf Y_{b,c}}(\vectsf y_{b,c})} \IEEEeqnarraynumspace
\end{IEEEeqnarray}
among all codewords
$\left \{ \vectsf s_{\e,2} (m',s),  \vectsf s_{\e,1} (m',s) \right \}$.


\section{Main Results}

For each $b \in [\eta]$, let $\gamma_{b,1} \ge \gamma_{b,2} \ge \ldots \ge \gamma_{b,q}$ be the $q$ largest eigenvalues of $\vectsf H_{b,s}^H \vectsf H_{b,s} $. We have the following lemma on the target detection probability.  
\begin{lemma} \label{lemma1}
The target detection probability is given by
\begin{IEEEeqnarray}{rCl}
P_{b, \D} = 1- F_{\tilde{\mathcal X}_2}  \left (F^{-1}_{\tilde{\mathcal X}_1} (1-P_{\text{FA}}) \right),
\end{IEEEeqnarray}
where $P_{\text{FA}}$ is the desired false alarm probability, $\tilde {\mathcal X}_1 (\{ w_{b,j}\}_{j =1}^{q}, \ell, \{ \nu_{b,j}\}_{j = 1}^q)$ and $\tilde {\mathcal X}_2(\{ w_{b,j}\}_{j =1}^{q}, \ell, \{\tilde \nu_{b,j}\}_{j = 1}^q)$ are generalized chi-square distributions with
\begin{subequations}\label{eq:27}
\begin{IEEEeqnarray}{rCl}
w_{b,j}&:=&\frac{\gamma_{b,j} (1-\kappa_1) \P}{1+\gamma_{b,j} (1-\kappa_1)\P},\\
 \nu_{b,j} &:=& \kappa_2 \ell \P \left (\frac{1-w_{b,j}}{w_{b,j}}\right)^2, \IEEEeqnarraynumspace\\
\tilde \nu_{b,j}&:=& \begin{cases}
\gamma_{b,j} \ell \P \left (\bsf + \frac{\kappa_2}{w_{b,j}}\right) & b \notin \Barr, \\
\gamma_{b,j} \ell \P \left (\bc + \frac{\kappa_2}{w_{b,j}}\right) & \text{o.w.}
\end{cases}
\end{IEEEeqnarray}
\end{subequations}
where
\begin{IEEEeqnarray}{rCl}
\kappa_1&:=& \begin{cases} \kappa_2(1-\alsf)^2, & b \notin \Barr \\
\kappa_2(1-\alc)^2(1-\alss)^2,  & \text{o.w.}
\end{cases}\\
\kappa_2&:=& \begin{cases} (1-\bsf) , & b \notin \Barr \\
(1-\bc)(1-\bss),  & \text{o.w.}
\end{cases}
\end{IEEEeqnarray}
%
and $F_{\tilde{\mathcal X}_1}(\cdot)$ and $F_{\tilde{\mathcal X}_2}(\cdot)$ are the corresponding CDFs. 
\end{lemma}
\begin{IEEEproof}
See  Appendix~\ref{App:A}.
\end{IEEEproof}
For each $b \in [\eta]$, let $\lambda_{b,1} \ge \lambda_{b,2} \ge \ldots \ge \lambda_{b,q}$ be the $q$ largest eigenvalues of $\vectsf H_{b,c}^H \vectsf H_{b,c} $. We have the following lemma on the URLLC decoding error probability.
\begin{lemma} \label{lemma2}
The average URLLC decoding error probability is upper bounded by
\begin{eqnarray}\label{eq:boundu}
\epsilon_{b,\U} \le  P_{b-1, \D} P_{\U,1}   + (1-P_{b-1, \D})P_{\U,2}
\end{eqnarray}
where 
\begin{eqnarray}
P_{\U,1}&: =& \left (\tepf+ (\tepf) ^{M_{\U} M_v}+\teps  \right), \label{eq:Pu1}\\
P_{\U,2}&: =& \left (1 - \left (1-\frac{\teps}{M_{\U}M_{v}}\right)^{M_{\U} M_v} \right),\label{eq:Pu2}
\end{eqnarray}
\begin{IEEEeqnarray}{rCl}
\tilde \epsilon_{\U,1} &: =& Q \left (\frac{-\log(M_{\U} M_v) + \ell \C_{\U} - K_{\U} \log (\ell)}{\sqrt{\ell \mathsf V_{\U}}} \right) + \frac{B}{\sqrt{\ell} }, \IEEEeqnarraynumspace \label{eq:tepf} \\
\tilde \epsilon_{\U,2} &:=& \frac{2}{\ell^{K_{\U} }} \left ( \frac{\log 2}{\sqrt{2\pi \ell}} + \frac{2B } {\sqrt{\ell}}\right),\label{eq:teps}
\end{IEEEeqnarray}
 for some $K_{\U}>0$ and $B >0$ and where $Q(\cdot)$ is the $Q$-function and
\begin{IEEEeqnarray}{rCl}
\C_{\U} : =  \sum_{j = 1}^q \C(\Omega_{b,j}),\quad 
\mathsf V_{\U} := \sum_{j = 1}^q \mathsf V(\Omega_{b,j}),
\end{IEEEeqnarray}
with $\C(x) = \frac{1}{2} \log (1+x)$, $\mathsf V(x): = \frac{x(2+x)}{2(1+x)^2}$ and 
\begin{equation}\label{eq:omegabj}
\Omega_{b,j}: = \frac{\lambda_{b,j} (1- (1-\bc)(1-\alc)^2)\P }{1+\lambda_{b,j}(1-\alc)^2(1-\bc)\P}.
\end{equation}
\end{lemma}
\begin{IEEEproof}
See  Appendix~\ref{App:B}. The proof follows by analyzing the following three error events: 
\begin{eqnarray}
\mathcal E_{\U,1}&:=& \{b \notin \Bdet | b \in \Barr \},\\
\mathcal E_{\U, 2}&:=& \{b \notin \Bdec | b \in \Bdet,  b \in \mathcal B_{\text{arrival}}\}, \\
\mathcal E_{\U,3}&:=& \{b \in \Bdet | b \notin \Barr \},
\end{eqnarray}
where $\mathcal E_{\U,1}$ happens when the transmitted URLLC message is not detected, $\mathcal E_{\U,2}$ happens when the transmitted URLLC message is detected by the UE but the decoded message does not match the transmitted one, and $\mathcal E_{\U,3}$ happens if no URLLC message has been transmitted but the UE incorrectly declares the detection of a URLLC message. 
\end{IEEEproof}
The following lemma is on the eMBB transmission rate.
\begin{lemma}\label{lemma3}
The eMBB transmission rate $\R_{\e}:= \frac{\log M_{\e}}{n}$ is upper bounded by
\begin{eqnarray}
\R_{\e} \le  {\C}_{\e} - \sqrt{ \frac{ \mathsf V_{\e}}{n}}Q^{-1}(\epsilon_{\e}-\Delta_{\e}) - K_{\e} \frac{\log (n)}{n} - \frac{\log(M_{\s})}{n}
\end{eqnarray}
for some $K_{\e} > 0$ and where 
\begin{IEEEeqnarray}{rCl}
\C_{\e} &:=&\sum_{\Bdt} \P_{\text{det}}^{|\Bdt|} (1-\P_{\text{det}})^{\eta-|\Bdt|} \tilde {\C}_{\e},\label{eq:Ce}\\
\mathsf V_{\e} &:=& \sum_{\Bdt} \P_{\text{det}}^{|\Bdt|} (1-\P_{\text{det}})^{\eta-|\Bdt|} \tilde {\mathsf V}_{\e},\label{eq:Ve}\\
\tilde{\C}_{\e}&:=&\frac{1}{\eta}\sum_{j = 1}^q \left (\sum_{b \notin \Bdt} \C(\Omega_{b,j}^{(1)}) + \sum_{b \in \Bdt} \C( \Omega_{b,j}^{(2)})\right),\label{eq:tce}\\
\tilde{\mathsf V}_{\e}&:=&\frac{1}{\eta} \sum_{j = 1}^q \left (\sum_{b \notin \Bdt} \mathsf V(\Omega_{b,j}^{(1)}) + \sum_{b \in \Bdt} \mathsf V(\Omega_{b,j}^{(2)})\right),\label{eq:tve}
\end{IEEEeqnarray}
where $\C(x) = \frac{1}{2} \log (1+x)$, $\mathsf V(x): = \frac{x(2+x)}{2(1+x)^2}$,
\begin{IEEEeqnarray}{rCl}
\Omega_{b,j}^{(1)}&: =& \frac{\lambda_{b,j}(1-(1-\alsf)^2(1-\bsf))\P}{1 + \lambda_{b,j} (1-\alsf)^2(1-\bsf) \P},\\
\Omega_{b,j}^{(2)} &:=& \frac{\lambda_{b,j}(1-\alc)^2(1-\bc)(1-(1-\alss)^2(1-\bss))\P}{1 + \lambda_{b,j} (1-(1-\alc)^2(1-\bc)) \P}, \notag \\ \\
\P_{\text{det}}&: =& P_{b-1,\D} (1- (\tepf)^{M_{\U}M_v}) +  (1-P_{b-1,\D}) P_{\U,2},\\
\Delta_{\e}&:=&\Bigg [ 1 + \frac{\tilde B}{\sqrt{n}}\left ( 1 + \frac{4}{n^K_{\e}}\right)  +\frac{2\log 2}{n^{K_{\e}}\sqrt{2n\pi}} \notag \\
&& \hspace{0.2cm}- \frac{(P_{b-1, \D} - \kappa_{\e})^{|\Bdt|} (( \tepf)^{M_{\U} M_v} - \kappa_{\e})^{\eta -|\Bdt|} }{\P_{\text{det}}^{|\Bdt|} (1-\P_{\text{det}} )^{\eta-|\Bdt|}}\Bigg], \label{eq:Deltae} \IEEEeqnarraynumspace
\end{IEEEeqnarray}
with $\kappa_{\e} : = P_{b-1, \D}(\tepf)^{M_{\U}M_v}$, for some $\tilde B > 0$, and where  $P_{\U,2}$ and $\tepf$  are defined in\eqref{eq:Pu2} and \eqref{eq:tepf}, respectively. 
\end{lemma}
\begin{IEEEproof}
See  Appendix~\ref{App:C}.
\end{IEEEproof}
By combining Lemmas~\ref{lemma1}--\ref{lemma3} with Proposition~\ref{prop1}, we have the following theorem on the rate-reliability-detection trade-off.
\begin{theorem}\label{th1}
Given $n$ and $\P$, the rate-reliability-detection trade-off is given by 
\begin{subequations}
\begin{IEEEeqnarray}{rCl}
\max_{\vect \beta, \vect \alpha} \;&& \C_{\e} - \sqrt{ \frac{\mathsf V_{\e}}{n}}Q^{-1}(\epsilon_{\e} - \Delta_{\e}) - K_{\e} \frac{\log (n)}{n} - \frac{\log(M_{\s})}{n}\label{eq:embb}\\ \notag \\
\text{s.t.:}\;  && P_{b-1, \D} P_{\U,1}   + (1-P_{b-1, \D})P_{\U,2} \le \epsilon_{\U}, \quad \forall b \in [\eta], \label{eq:urllc}\IEEEeqnarraynumspace\\ \notag \\
&& 1- F_{\tilde{\mathcal X_2}} (F^{-1}_{\tilde{\mathcal X_1}} (1-P_{\text{FA}})) \ge \P_{\D}, \quad \forall b \in [\eta ],
\end{IEEEeqnarray}
\end{subequations}
where $\vect \beta:=\{\bc, \bsf,\bss\}$ and $\vect \alpha:=\{\alc,\alsf,\alss\}$.  
\end{theorem}
\begin{figure}[t]
\center
				\includegraphics[width=0.37\textwidth]{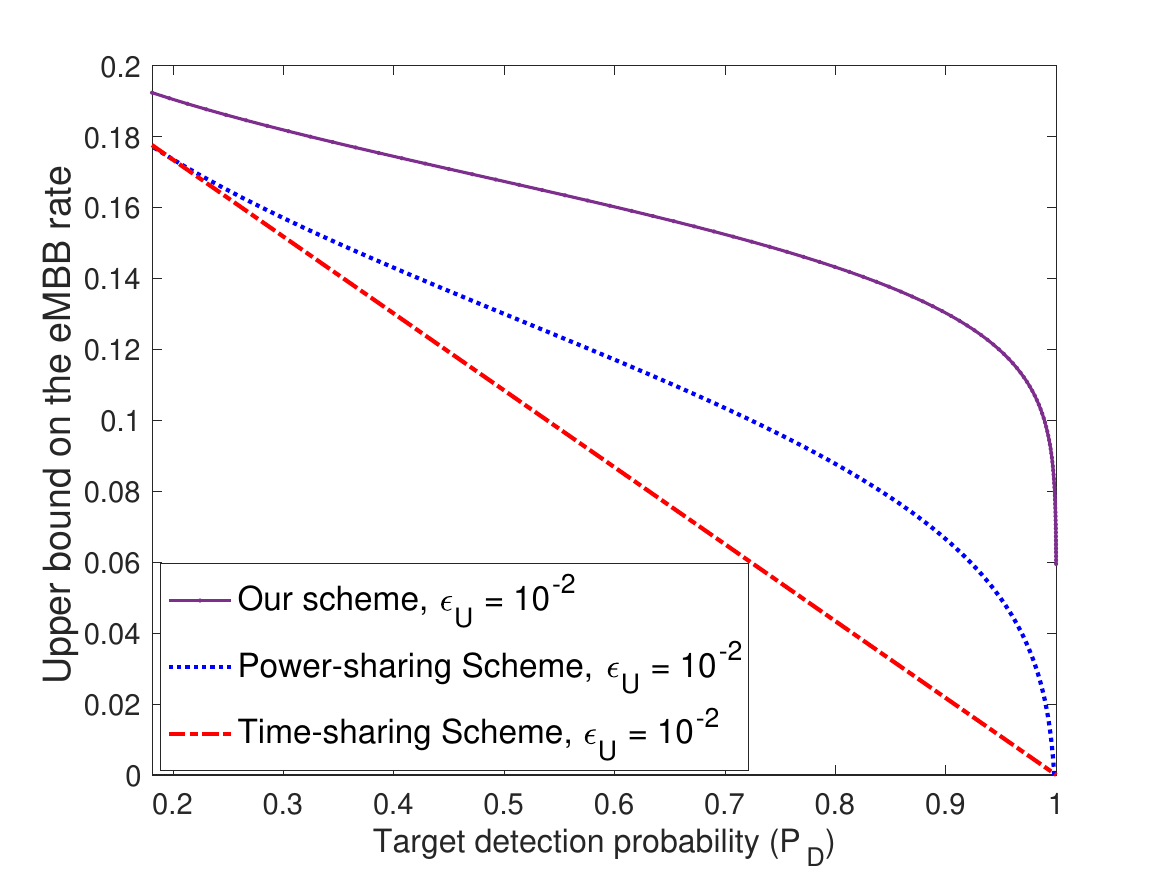}
\caption{Comparing our DPC-based scheme with power-sharing and time-sharing schemes for $\P= 0.5$, $\ell = 150$, $\eta = 10$, $P_{FA} = 10^{-6}$, $\epsilon_e = 10^{-3}$. }
\label{fig2}
\vspace{-0.4cm}
\end{figure}

\begin{figure}[t]
\center
				\includegraphics[width=0.37\textwidth]{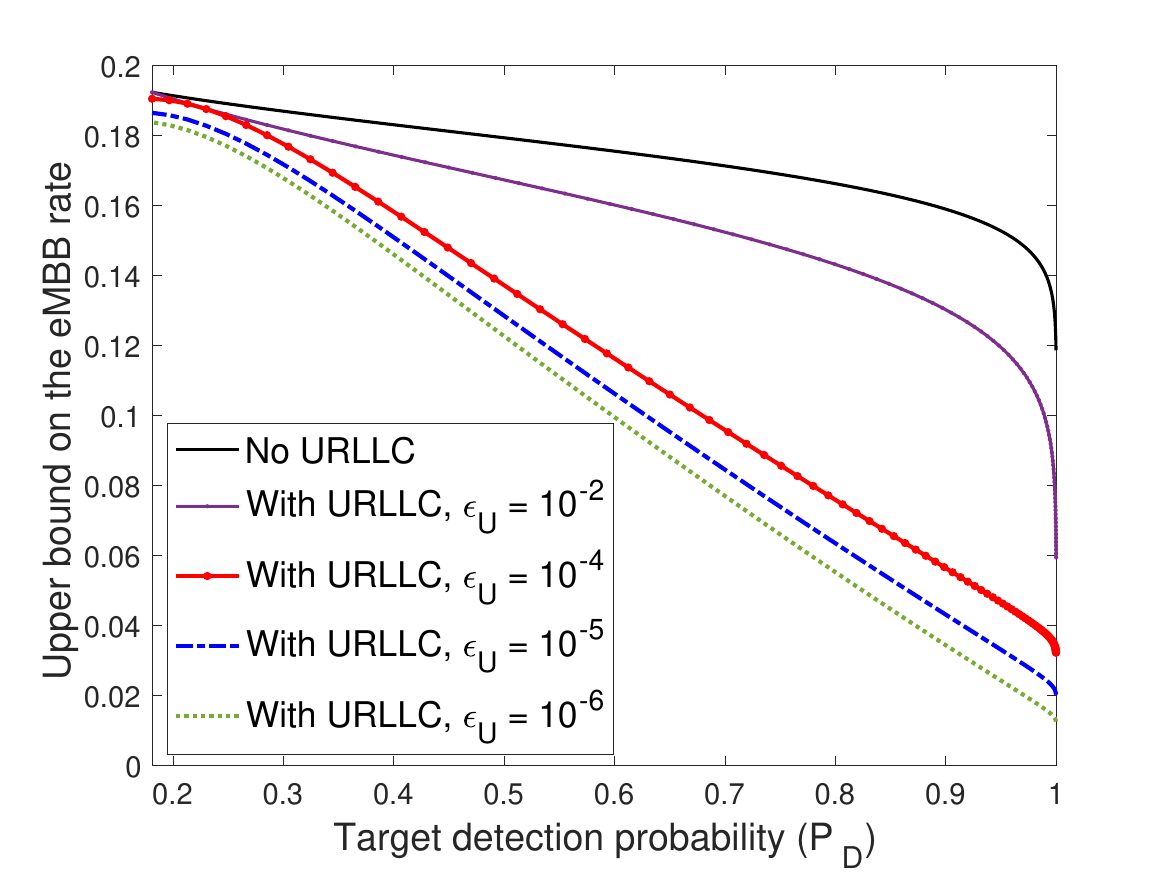}
\caption{Our proposed  scheme with  different values of the URLLC decoding error probability threshold for $\P= 0.5$, $\ell = 150$, $\eta = 10$, $P_{FA} = 10^{-6}$, $\epsilon_e = 10^{-3}$.}
\label{fig3}
\vspace{-0.4cm}
\end{figure}

\section{Numerical Analysis}
In this section, we numerically evaluate the rate-reliability-detection trade-off of Theorem~\ref{th1}. 
 Given $\P, n$ and $\eta$ and $\epsilon_{\U}$, we first find values of $\alc$ and $\bc$ such that $\epsilon_{b, \U}$ is below  $\epsilon_{\U}$  in all blocks. Meaning that the constraint \eqref{eq:urllc} is satisfied for all blocks. Next, we maximize the upper bound on the eMBB rate  in \eqref{eq:embb} over $\bsf, \bss, \alsf, \alss$ while satisfying the minimum required target detection probability (i.e., $\P_{\D}$) over all blocks. 

 Fig.~\ref{fig2} illustrates the upper bound on the eMBB transmission rate as a function of $\P_{\D}$ for  our DPC-based scheme, power-sharing and  time-sharing schemes when $\epsilon_{\U}$ is fixed at $10^{-2}$. In the power-sharing scheme the BS sends a linear combination of  the sensing and communication signals and shares the transmit power $\P$ between the two tasks. In the time-sharing scheme sensing and communication tasks are performed independently each using a portion of the blocklength $n$. As can be seen from this figure,  our scheme significantly outperforms the other two schemes. 
 Fig.~\ref{fig3} illustrates the upper bound on the eMBB transmission rate as a function of $\P_{\D}$ for our  proposed scheme under different  levels of  URLLC reliability. It can be seen that as we decrease the required threshold on the URLLC decoding error probability (i.e., $\epsilon_{\U}$) the eMBB rate decreases.    

\section{Conclusion}
We proposed a MIMO  ISAC-enabled URLLC system where a BS communicates with a UE and a SR collects echo signals reflected from a target of interest.  The BS simultaneously transmit messages from eMBB and URLLC services. During each eMBB transmission, the transmission of an additional URLLC message is triggered when the SR sensed the presence of the target. To reinforce URLLC transmissions, the interference of both eMBB and sensing signals   were mitigated using DPC. For this system, we  formulated the rate-reliability-detection trade-off in the finite blocklength regime.  Our numerical analysis shows a significant outperformance of our scheme over  the power-sharing  and  the  time-sharing methods. 
\section*{Acknowledgement}
This work was supported in part by NextG Innovation Award from Princeton University.

\appendices 
\section{Proof of Lemma~\ref{lemma1}} \label{App:A}
The SR treats the interference of the communication signal as noise. At each block~$b \notin \Barr$, the communication signal only carries eMBB traffic, i.e.,  $\vectsf X_{b,c} = \vectsf X_{b}^{(\e,1)}$, whereas, at each block $b \in \Barr$, the communication signal carries both eMBB  and URLLC traffic, i.e.,  $\vectsf X_{b,c} = \vectsf X_b^{(\U)} + \vectsf X_{b}^{(\e,2)}$.
Denote by $q: = \min \{t,r\}$ and let $\gamma_{b,1} \ge \gamma_{b,2} \ge \ldots \ge \gamma_{b,q}$ be the $q$ largest eigenvalues of $\vectsf H_{b,s}^H \vectsf H_{b,s} $.  The LRT test in each block~$b \in [\eta]$  is given by 
\begin{IEEEeqnarray}{rCl}
T_b &=& \log  \left ( \prod_{j = 1}^{q} \frac{f\left (\vect Y_{b,s,j}  | \mathcal H_1\right)}{f\left(\vect Y_{b,s,j}  | \mathcal H_0\right)} \right) \underset{{\mathcal H_0}}{\overset{\mathcal H_{1}}{\gtrless}} \delta.
\end{IEEEeqnarray} 
Given that $\vect N_{b,s,j} \sim \mathcal N(0, \vect I_{\ell})$, thus $f\left(\vect Y_{b,s,j}  | \mathcal H_0\right)$ is an i.i.d Gaussian distribution. However, due to our power-shell code construction, $f\left(\vect Y_{b,s,j}  | \mathcal H_1\right)$ is non i.i.d. Hence, we take a change of metric measurement approach by introducing the following new LRT test: 
\begin{equation}
\tilde T_b := \log  \left ( \prod_{j = 1}^{q} \frac{Q\left (\vect Y_{b,s,j}  | \mathcal H_1\right)}{f\left(\vect Y_{b,s,j}  | \mathcal H_0\right)} \right) \underset{{\mathcal H_0}}{\overset{\mathcal H_{1}}{\gtrless}} \delta,
\end{equation}
where $Q\left(\vect Y_{b,s,j}  | \mathcal H_1\right)$ is i.i.d Gaussian distribution  $\mathcal N(\vect \mu_{b,j}, \sigbj \vect I_{\ell})$. 
If $b \notin \Barr$,  $\vect \mu_{b,j} = \sqrt{\gamma_{b,j}}\vect x_{b,j}^{(\s,1)}$ and $\sigbj = 1 + \gamma_{b,j}(1-(1-\bsf)(1-\alsf)^2)\P$. Otherwise, $\vect \mu_{b,j} = \sqrt{\gamma_{b,j}}\vect x_{b,j}^{(\s,2)}$ and $\sigbj = 1+\gamma_{b,j} (1- (1-\bc)(1-\alc)^2(1-\bss)(1-\alss)^2)\P$. 
The test thus can be written as
\begin{IEEEeqnarray}{rCl}
\tilde T_b = \sum_{j = 1}^q \left (|| \vect Y_{b,s,j}||^2 -\frac{ ||\vect Y_{b,s,j} -  \vect \mu_{b,j} ||^2}{\sigbj}\right)\underset{{\mathcal H_0}}{\overset{\mathcal H_{1}}{\gtrless}} \delta. \IEEEeqnarraynumspace
\end{IEEEeqnarray}
The probability of false alarm is given by
\begin{IEEEeqnarray}{rCl}
P_{b, \text{FA}} &=& \Pr [\tilde T_b > \delta | \mathcal H_0] \\
& = & \Pr \left [ \sum_{j = 1}^q \left (|| \vect N_{b,s,j}||^2 - \frac{||\vect N_{b,s,j} -  \vect \mu_{b,j}  ||^2}{\sigbj} \right) > \delta \right ] \IEEEeqnarraynumspace\\
&=&\Pr \Big [ \sum_{j = 1}^q (1-\frac{1}{\sigbj}) ||\vect N_{b,s,j}+ \frac{1}{\sigbj -1}\vect \mu_{b,j} ||^2 \notag \\
&& \hspace{2.2cm}> \delta +  \sum_{j = 1}^q\frac{1}{\sigbj-1}||\vect \mu_{b,j}||^2  \Big ] \\
& = & \Pr \left [ u_b  > \delta +  \sum_{j = 1}^q\frac{1}{\sigbj-1}||\vect \mu_{b,j}||^2   \right ],
\end{IEEEeqnarray}
where 
\begin{equation}
u_b := \sum_{j = 1}^q (1-\frac{1}{\sigbj}) ||\vect N_{b,s,j}+ \frac{1}{\sigbj -1}\vect \mu_{b,j} ||^2.
\end{equation}
Given that $\vect N_{b,s,j} \sim  \mathcal N(\vect 0, \I_{\ell})$, then   $\vect N_{b,s,j} + \frac{1}{\sigbj -1}\vect \mu_{b,j} \sim \mathcal N( \frac{1}{\sigbj -1}\vect \mu_{b,j}, \vectsf I_{\ell})$ and consequently $||\vect N_{b,s,j}+ \frac{1}{\sigbj -1}\vect \mu_{b,j} ||^2
$ follows a non-central chi-square distribution with degree of $\ell$ and parameter $\frac{1}{(\sigbj -1)^2}||\vect \mu_{b,j}||^2$ . Hence, $u_{b}$ is the weighted sum of independent non-central chi-square random variables. It thus follows a generalized chi-square distribution. More specifically, 
\begin{equation}
u_b \sim \tilde {\mathcal X}_1 (\{ w_{b,j}\}_{j =1}^{q}, \ell, \{ \nu_{b,j}\}_{j = 1}^q),
\end{equation}
where the parameters $\{ w_{b,j}\}_{j =1}^{q}$ and $\{ \nu_{b,j}\}_{j = 1}^q$ are defined in \eqref{eq:27}. 
Let $F_{\tilde{\mathcal X}_1} (\cdot)$ denote the CDF of the corresponding generalized chi-square distribution.  The probability of false alarm thus is given by
\begin{eqnarray}
P_{b, \text{FA}} = 1 - F_{\tilde{\mathcal X}_1}\left (\delta +  \sum_{j = 1}^q\frac{1}{\sigbj-1}||\vect \mu_{b,j}||^2\right).
\end{eqnarray}
For each block~$b$, we fix the false alarm probability at $P_{\text{FA}}$. The threshold $\delta$ then is equal to
\begin{eqnarray}
\delta = F^{-1}_{\tilde{\mathcal X}_1} (1-P_{\text{FA}})-\sum_{j = 1}^q\frac{1}{\sigbj-1}||\vect \mu_{b,j}||^2.
\end{eqnarray} 
The target detection probability is given by
\begin{IEEEeqnarray}{rCl}
\lefteqn{P_{b, \D}} \notag \\
 &=& \Pr [\tilde T_b > \delta | \mathcal H_1] \\
& = & \Pr \Big [ \sum_{j = 1}^q \Big (|| \vect N_{b,s,j}  +  \vect X_{b,j} \sqrt{\gamma_{b,j}} ||^2  \notag \\
&& \hspace{1.5cm}-\frac{ ||\vect N_{b,s,j} + \vect X_{b,c,j} \sqrt{\gamma_{b,j}} ||^2}{\sigbj}\Big ) > \delta \Big ] \\
&=&\Pr \Big [ \sum_{j = 1}^q \Big ((1-\frac{1}{\sigbj}) ||\vect N_{b,s,j} + \vect X_{b,c,j} \sqrt{\gamma_{b,j}} \notag \\
&& \hspace{0.7cm}+ \frac{\sigbj}{\sigbj-1} \vect \mu_{b,j}||^2 \Big) > \delta +  \sum_{j = 1}^q\frac{1}{\sigbj-1}||\vect \mu_{b,j}||^2 \Big] \IEEEeqnarraynumspace\\
& = & \Pr \left [ \tilde u_b  > F^{-1}_{\tilde{\mathcal X}_1}(1-P_{\text{FA}}) \right ],
\end{IEEEeqnarray}
where
\begin{equation}
\tilde u_b : = \sum_{j = 1}^q (1-\frac{1}{\sigbj}) ||\vect N_{b,s,j} +\sqrt{\gamma_{b,j}}  \vect X_{b,c,j} + \frac{\sigbj}{\sigbj-1} \vect \mu_{b,j}||^2 
\end{equation}
Following the same argument provided for the false alarm probability, one can obtain that $\tilde u_b$ is the weighted sum of independent non-central chi-square random variables and thus follows a generalized chi-square distribution as 
\begin{IEEEeqnarray}{rCl}
\tilde u_b &\sim&  \tilde {\mathcal X}_2(\{ w_{b,j}\}_{j =1}^{q}, \ell, \{\tilde \nu_{b,j}\}_{j = 1}^q),
\end{IEEEeqnarray}
where the parameters $\{ w_{b,j}\}_{j =1}^{q}$ and $\{\tilde \nu_{b,j}\}_{j = 1}^q$ are defined in \eqref{eq:27}. 
Let $F_{\tilde{\mathcal X}_2} (\cdot)$ denote the CDF of the corresponding generalized chi-square distribution. 

Subsequently, the target detection probability is given by 
\begin{eqnarray}
P_{b, \D} = 1- F_{\tilde{\mathcal X}_2} \left (F^{-1}_{\tilde{\mathcal X}_1} (1-P_{\text{FA}})\right).
\end{eqnarray} 
This concludes the proof.

\section{Proof of Lemma~\ref{lemma2}}\label{App:B}
When channel state is known at both the BS and the UE, then by performing a singular value decomposition,  the MIMO channel in \eqref{eq:channel} is transferred into the following set of $q:=\min \{t,r\}$ parallel channels 
\begin{IEEEeqnarray}{rCl}
\vect Y_{b,c,j} = \vect X_{b,j} \sqrt{\lambda_{b,j}} + \vect N_{b,c,j},
\end{IEEEeqnarray}
for each $j \in [q]$ and each $b \in [\eta]$. Here, $\lambda_{b,1} \ge \lambda_{b,2} \ge \ldots \ge \lambda_{b,q}$ are the largest $q$ eigenvalues of $\vectsf H_{b,c} \vectsf H_{b,c}^H$ and $\vect N_{b,c,j} \sim \mathcal{N}(\vect 0, \vectsf I_{\ell})$ are independent noise vectors.  Let $\vect \lambda_b = [\lambda_{b,1}, \ldots, \lambda_{b,q}]$. We further assume that the encoders $f_{\e}$ and $f_{b}$ in Definition~\ref{def:code} act on $\vect \lambda_b$ instead of $\vectsf H_{b,c}$.
Also, let  $\vectsf X_b : = [\vect X_{b,1}, \ldots, \vect X_{b,q}]$. The power constraint \eqref{eq:1} applies that
\begin{equation}
\sum_{j = 1}^q ||\vect X_{b,j}||^2 \le \ell \P. 
\end{equation} 
Note that, to improve the finite blocklength performance, all the codewords are uniformly distributed on the power shell. According to Shannon’s observation, the optimal decay of the probability of error near capacity of the point-to-point Gaussian channel is achieved by codewords on the power-shell \cite{Shannon1959}. As a result of this code construction, the induced output distributions $f_{\vectsf Y_{b,c}| \vectsf V_{b}}(\vectsf y_{b,c}| \vectsf v_{b})$ and $f_{\vectsf Y_{b,c}}(\vectsf y_{b,c})$
are non i.i.d. We thus take a change of metric approach and instead of the information density metric $i^{(\U)}_b  (\vectsf V_b; \vectsf Y_{b,c} )$ defined in \eqref{eq:ibu}, we use the following metric: 
\begin{equation}
\tilde i^{(\U)}_b  (\vectsf V_b; \vectsf Y_{b,c} ):= \log \frac{Q_{\vectsf Y_{b,c}| \vectsf V_{b}} (\vectsf y_{b,c}| \vectsf v_{b})}{Q_{\vectsf Y_{b,c}}(\vectsf y_{b,c})}. 
\end{equation}
where $Q$'s are i.i.d Gaussian distributions. 

Given that the channel state information is known at both BS and UE, the information density metric can be written as 
\begin{equation}
\tilde i^{(\U)}_b  (\vectsf V_b; \vectsf Y_{b,c} ) = \sum_{j = 1}^q \tilde i^{(\U)}_b  (\vbj; \vect Y_{b,c,j} ).
\end{equation}
The following lemma shows that the random variable $\tilde i^{(\U)}_b  (\vectsf V_b; \vectsf Y_{b,c} )$ converges in distribution to a Gaussian distribution. 
\begin{lemma}\label{lemma4}
The following holds: 
\begin{IEEEeqnarray}{rCl}
\tilde i^{(\U)}_b  (\vectsf V_b; \vectsf Y_{b,c} ) \sim \mathcal N(\ell \C_{\U},\ell \mathsf V_{\U} ),
\end{IEEEeqnarray}
where 
\begin{IEEEeqnarray}{rCl}
\C_{\U} : =  \sum_{j = 1}^q \C(\Omega_{b,j}),\quad 
\mathsf V_{\U} := \sum_{j = 1}^q \mathsf V(\Omega_{b,j})
\end{IEEEeqnarray}
with $\C(x) = \frac{1}{2} \log (1+x)$, $\mathsf V(x): = \frac{x(2+x)}{2(1+x)^2}$ and $\Omega_{b,j}$ is defined in \eqref{eq:omegabj}. 
\begin{IEEEproof}
The proof follows the same argument as the proof of \cite[Proposition~1]{MolavianJaziArXiv}.
\end{IEEEproof}
\end{lemma}
We are now ready to analyze the URLLC decoding error probability. Recall the definition of $\Barr, \Bdet$ and $\Bdec$ from \eqref{eq:barr}, \eqref{eq:bdet} and \eqref{eq:bdec}, respectively.  
In each block~$b \in \Barr$, we have the following URLLC decoding error events:
\begin{eqnarray}
\mathcal E_{\U,1}&:=& \{b \notin \Bdet | b \in \Barr \},\\
\mathcal E_{\U, 2}&:=& \{b \notin \Bdec | b \in \Bdet,  b \in \mathcal B_{\text{arrival}}\}. 
\end{eqnarray}
In each block $b \notin \Barr$ we have the following error event:
\begin{eqnarray}
\mathcal E_{\U,3}&:=& \{b \in \Bdet | b \notin \Barr \}.
\end{eqnarray}
The URLLC decoding error probability thus can be bounded by 
\begin{eqnarray}
\epsilon_{b,\U} \le P_{b-1, \D} \left (\Pr [\mathcal E_{\U,1}] + \Pr [\mathcal E_{\U,2}] \right)+ (1- P_{b-1, \D} ) \Pr [\mathcal E_{\U,3}]. \label{eq:57}
\end{eqnarray}
In the following sections, we evaluate the occurrence probability of each event individually.  
\subsubsection{Analyzing $\Pr [\mathcal E_{\U,1}]$} This error event is equivalent to the probability that for all $v \in [ M_{v}]$ and for all $m \in [ \ldots, M_{\U}]$ there is no codeword $ \vectsf V_b(m,v)$ such that $\tilde i^{(\U)}_b  (\vectsf V_b; \vectsf Y_{b,c} ) > \delta_{\U}$. Hence, 
\begin{eqnarray}
\Pr [\mathcal E_{\U,1}] = \left (\Pr [ \tilde i^{(\U)}_b  (\vectsf V_b; \vectsf Y_{b,c} ) \le \delta_{\U} ] \right)^{M_{\U} M_{v}}. \label{eq:58}
\end{eqnarray}
\subsubsection{Analyzing $\Pr [\mathcal E_{\U,2}]$}
This error event is equivalent to the missed detection of the transmitted URLLC message. To evaluate this probability, we use the threshold bound for maximum-metric decoding. For any given threshold $\delta_{\U}$: 
\begin{IEEEeqnarray}{rCl}
\Pr [\mathcal E_{\U,2}] &\le& \Pr [\tilde i^{(\U)}_b  (\vectsf V_b(m_{b,\U}, v); \vectsf Y_{b,c} ) \le \delta_{\U} ] \notag \\
&& + (M_{\U} M_v -1)  \Pr [\tilde i^{(\U)}_b  (\bar {\vectsf V}_b(m', v'); \vectsf Y_{b,c} ) > \delta_{\U} ] \label{eq:59} \IEEEeqnarraynumspace
\end{IEEEeqnarray}
where $m' \in [M_{\U}]$, $v' \in [M_v]$ and $(m_{b,\U}, v) \neq (m', v')$ and $\bar {\vectsf V}_b \sim f_{\vectsf V_b}$ and is independent of $\vectsf V_b$ and $ \vectsf Y_{b,c} $. 

\subsubsection{Analyzing $\Pr [\mathcal E_{\U,3}]$} This error event is equivalent to the probability that no URLLC has been sent over block~$b$ but there is at least one codeword $\bar {\vectsf V}_b$ such that $\Pr [\tilde i^{(\U)}_b  (\bar{\vectsf V}_b; \vectsf Y_{b,c} ) > \delta_{\U}]$. Hence,
\begin{eqnarray}
\Pr [\mathcal E_{\U,3}] = 1 - \left (\Pr [\tilde i^{(\U)}_b  (\bar{\vectsf V}_b; \vectsf Y_{b,c} ) \le \delta_{\U}]\right )^{M_{\U} M_{v}} \label{eq:60}
\end{eqnarray}
Combining \eqref{eq:57}, \eqref{eq:58}, \eqref{eq:59} and \eqref{eq:60}, we have
\begin{IEEEeqnarray}{rCl}
\epsilon_{b,\U}
&\le& P_{b-1, \D} \Big( \Pr [ \tilde i^{(\U)}_b  (\vectsf V_b; \vectsf Y_{b,c} ) \le \delta_{\U} ]^{M_{\U} M_{v}} \notag \\
&&\hspace{1.25cm}+ \Pr [\tilde i^{(\U)}_b  (\vectsf V_b; \vectsf Y_{b,c} ) \le \delta_{\U} ] \notag \\
&&\hspace{1.5cm}+ (M_{\U} M_v -1)  \Pr [\tilde i^{(\U)}_b  (\bar {\vectsf V}_b; \vectsf Y_{b,c} ) > \delta_{\U} ]\Big)\notag \\
&&\hspace{-0.7cm}+ (1- P_{b-1, \D} )\left(1 - \left (\Pr [\tilde i^{(\U)}_b  (\bar{\vectsf V}_b; \vectsf Y_{b,c} ) \le \delta_{\U}]\right )^{M_{\U} M_{v}}\right) . \label{eq:euf}\IEEEeqnarraynumspace
\end{IEEEeqnarray}

set 
\begin{eqnarray}
\delta_{\U} := \log(M_{\U}M_v)+ K_{\U} \log (\ell) 
\end{eqnarray}
for some $K_{\U} > 0$. 
To evaluate  $\Pr [\tilde i^{(\U)}_b  (\bar {\vectsf V}_b; \vectsf Y_{b,c} ) > \delta_{\U} ]$ we follow \cite[Lemma 47]{Yuri2012} which results in 
\begin{eqnarray}
\Pr [\tilde i^{(\U)}_b  (\bar {\vectsf V}_b; \vectsf Y_{b,c} ) > \delta_{\U} ] \le \frac{2}{M_{\U} M_{v} \ell^{K_{\U} }} \left ( \frac{\log 2}{\sqrt{2\pi\ell}} + \frac{2B} {\sqrt{\ell}}\right),
\end{eqnarray}

for some $B > 0$. To evaluate $\Pr [\tilde i^{(\U)}_b  (\vectsf V_b(m_{b,\U}, v); \vectsf Y_{b,c} ) < \delta_{\U}]$, we employ  the Berry-Esseen central limit theorem (CLT) for functions \cite[Proposition~1]{MolavianJaziArXiv} which results in 
\begin{IEEEeqnarray}{rCl}
\lefteqn{\Pr [\tilde i^{(\U)}_b  (\vectsf V_b; \vectsf Y_{b,c} ) \le \delta_{\U}  ]} \notag \\
& \le& Q \left (\frac{-\log(M_{\U} M_v) + \ell \C_{\U} - K_{\U} \log (\ell)}{\sqrt{\ell \mathsf V_{\U}}} \right) + \frac{B}{\sqrt{\ell} }, \IEEEeqnarraynumspace
\end{IEEEeqnarray}
where $Q(\cdot)$ is the Q-function.
Denote by \begin{IEEEeqnarray}{rCl}
P_{\U,1}&: =& \left (\tepf+ (\tepf) ^{M_{\U} M_v}+\teps  \right), \label{eq:pu1}\\
P_{\U,2}&: =& \left (1 - \left (1-\frac{\teps}{M_{\U}M_{v}}\right)^{M_{\U} M_v} \right), \label{eq:pu2}
\end{IEEEeqnarray}
with
\begin{IEEEeqnarray}{rCl}
\tilde \epsilon_{\U,1} &: =& Q \left (\frac{-\log(M_{\U} M_v) + \ell \C_{\U} - K_{\U} \log (\ell)}{\sqrt{\ell \mathsf V_{\U}}} \right) + \frac{B}{\sqrt{\ell }},\label{eq:ep1} \IEEEeqnarraynumspace \\
\tilde \epsilon_{\U,2} &:=& \frac{2}{ \ell^{K_{\U}}} \left ( \frac{\log 2}{\sqrt{2\pi \ell}} + \frac{2B} {\sqrt{\ell }}\right).\label{eq:ep2}
\end{IEEEeqnarray}
Combining \eqref{eq:pu1}--\eqref{eq:ep2} with \eqref{eq:euf} concludes the proof. 

\section{Proof of Lemma~\ref{lemma3}} \label{App:C}
The eMBB message is decoded at the end of the entire $n$ channel uses. We have the following two eMBB decoding error events:
\begin{eqnarray}
\mathcal E_{\e,1} &: =& \{\Bdet \neq \Barr\} \\
\mathcal E_{\e,2} &: =& \{\hat m_{\e} \neq m_e | \Bdet = \Barr\} 
\end{eqnarray}
 Recall the definition of $B_{\text{dt}}$ as  the realization of the set $ \mathcal B_{\text{detect}}$. The average eMBB decoding error probability is bounded by
\begin{IEEEeqnarray}{rCl}
\epsilon_{\e}^{(n)} &\le&
\sum_{B_{\text{dt}}} \Pr [\Bdet = B_{\text{dt}}] \notag \\
&&\Big ( \Pr [\mathcal E_{\e,1}| \Bdet = B_{\text{dt}}] + \Pr [\mathcal E_{\e,2} | \Bdet = B_{\text{dt}}] \Big). \label{eq:71}
\end{IEEEeqnarray}
\subsubsection{Analyzing $\Pr [\Bdet = B_{\text{dt}}] $} Define
\begin{IEEEeqnarray}{rCl}
\P_{\text{det}} &: =& \Pr [b \in \Barr] \Pr [b \in \Bdet | b \in \Barr] \notag \\
&& + \Pr [b \notin \Barr] \Pr [b \in \Bdet | b \notin \Barr] \\
& = & P_{b-1,\D} (1- \Pr [\mathcal E_{\U,1}]) +  (1-P_{b-1,\D}) \Pr [\mathcal E_{\U,3}] \\
& = & P_{b-1,\D} (1- (\tepf)^{M_{\U}M_v}) \notag \\
&& +  (1-P_{b-1,\D}) \left(1-(1-\teps)^{M_{\U}M_v} \right) 
\end{IEEEeqnarray}
We thus have
\begin{eqnarray}
\Pr [\Bdet = B_{\text{dt}}]  = \P_{\text{det}}^{|\Bdt|} (1-\P_{\text{det}} )^{\eta-|\Bdt|}. \label{eq:74}
\end{eqnarray}

\subsubsection{Analyzing $\Pr [\mathcal E_{\e,1}| \Bdet =  B_{\text{dt}} ] $}
This error event is equivalent to the probability that the set of blocks in which a URLLC message is detected by the UE differs from the set of blocks in which URLLC messages have been transmitted by the BS.  
%
\begin{IEEEeqnarray}{rCl}
\lefteqn{\Pr [\mathcal E_{\e,1} | \Bdet =  B_{\text{dt}}]} \notag \\
 &=& \Pr [ \Bdet \neq \Barr | \Bdet = B_{\text{dt}} ] \\
& = & \Pr [ \Barr \neq B_{\text{dt}} | \Bdet = B_{\text{dt}} ] \\
&=&  1 - \Pr [  \Barr =  B_{\text{dt}}| \Bdet = B_{\text{dt}}  ] \\
& = & 1 - \frac{\Pr [\Barr =  B_{\text{dt}}] \Pr [ \Bdet =  B_{\text{dt}}| \Barr = B_{\text{dt}}  ]}{\Pr[\Bdet  = B_{\text{dt}}]} \\
& = & 1 - \frac{(P_{b-1, \D}(1- \Pr [\mathcal E_{\U,1}]))^{|\Bdt|} (\Pr [\mathcal E_{\U,1}](1- P_{b-1, \D}))^{\eta - |\Bdt|} }{\P_{\text{det}}^{|\Bdt|} (1-\P_{\text{det}} )^{\eta-|\Bdt|}}.\notag \\ \label{eq:79}
\end{IEEEeqnarray}

\subsubsection{Analyzing $\Pr [\mathcal E_{\e,2}|\Bdet =  B_{\text{dt}} ]$}
 To evaluate this probability, we first follow the argument provided in Appendix~\ref{App:B}, and introduce a new metric:
 \begin{IEEEeqnarray}{rCl}
\lefteqn{\tilde i^{(\e)} \left ( \vectsf s_{\e,2} ,  \vectsf s_{\e,1} ;  \vectsf y_c| \mathcal B_{\text{detect}} = B_{\text{dt}} \right):=}\notag \\ &&
 \log \hspace{-0.15cm}\prod_{b\notin  B_{\text{dt}}  }\hspace{0cm}  \frac{f_{\vectsf Y_{b,c}| \vectsf S_{b}^{(1)}} (\vectsf y_{b,c}| \vectsf s_{b}^{(1)})}{Q_{\vectsf Y_{b,c}}(\vectsf y_{b,c})} +  \log \hspace{-0.15cm}\prod_{b\in  B_{\text{dt}} } \hspace{0cm}  \frac{Q_{\vectsf Y_{b,c} | \vectsf S_{b}^{(2)}} (\vectsf y_{b,c}| \vectsf s_{b}^{(2)})}{Q_{\vectsf Y_{b,c}}(\vectsf y_{b,c})} \IEEEeqnarraynumspace
\end{IEEEeqnarray}
where $Q$s are i.i.d Gaussian distributions. Following the same argument as Lemma~\ref{lemma4}, we have  the following lemma showing $\tilde i^{(\e)} \left ( \vectsf s_{\e,2} ,  \vectsf s_{\e,1} ;  \vectsf y_c| \mathcal B_{\text{detect}} = B_{\text{dt}} \right)$ converges in distribution to a Gaussian distribution. 
\begin{lemma}
The following holds:
\begin{IEEEeqnarray}{rCl}
\tilde i^{(\e)} \left ( \vectsf s_{\e,2} ,  \vectsf s_{\e,1} ;  \vectsf y_c| \mathcal B_{\text{detect}} = B_{\text{dt}} \right) \sim  \mathcal N(n \tilde{\C}_{\e},n \tilde{\mathsf V}_{\e}) 
\end{IEEEeqnarray}
where $\tilde{\C}_{\e}$ and $\tilde{\mathsf V}_{\e}$ are defined in \eqref{eq:tce} and \eqref{eq:tve}, respectively. 
\end{lemma}
We now use the threshold bound for maximum-metric decoding to evaluate $\Pr [\mathcal E_{\e,2}|\Bdet =  B_{\text{dt}} ]$. For any given threshold $\delta_{\e}$: 
\begin{IEEEeqnarray}{rCl}
\lefteqn{\Pr [\mathcal E_{\e,2}|\Bdet = B_{\text{dt}} ]} \notag \\ 
&\le& \Pr [\tilde i^{(\e)} \left ( \vectsf S_{\e,1},  \vectsf S_{\e,2};  \vectsf Y_c| \mathcal B_{\text{detect}} = B_{\text{dt}} \right) \le \delta_{\e} ] \notag \\
&&+ (M_{\e}M_{\s} -1)  \Pr [\tilde i^{(\e)} \left ( \bar{ \vectsf S}_{\e,1},  \bar{ \vectsf S}_{\e,2};  \vectsf Y_c| \mathcal B_{\text{detect}} = B_{\text{dt}} \right) > \delta_{\e} ], \IEEEeqnarraynumspace \label{eq:80}
\end{IEEEeqnarray}
where $\bar{ \vectsf S}_{\e,1} \sim f_{ \vectsf S_{\e,1}}$ and $\bar{ \vectsf S}_{\e,2} \sim f_{ \vectsf S_{\e,2}}$ and are independent of $ \vectsf S_{\e,1}$, $ \vectsf S_{\e,2}$ and $ \vectsf Y_{c} $. 

Set
\begin{eqnarray}
\delta_{\e} := \log M_{\e} + \log M_{\s} + K_{\e} \log (n)
\end{eqnarray}
for some $K_{\e} > 0$.

To evaluate $\Pr [\tilde i^{(\e)} \left (  \vectsf S_{\e,1},  \vectsf S_{\e,2};  \vectsf Y_c| \mathcal B_{\text{detect}} = B_{\text{dt}} \right) \le \delta_{\e} ]$ we employ  the Berry-Esseen central limit theorem (CLT) for functions \cite[Proposition~1]{MolavianJaziArXiv} which results in
\begin{IEEEeqnarray}{rCl}
\lefteqn{\Pr [\tilde i^{(\e)} \left (  \vectsf S_{\e,1},   \vectsf S_{\e,2};  \vectsf Y_c| \mathcal B_{\text{detect}} = B_{\text{dt}} \right) \le \delta_{\e} ]} \notag \\
& \le& Q\left (\frac{-\log(M_{\e}) - \log(M_{\s}) + n \tilde {\C}_{\e} - K_{\e} \log (n)}{\sqrt{n \tilde {\mathsf V}_{\e}}} \right) + \frac{\tilde B}{\sqrt{n } }, \notag \\\label{eq:82}
\end{IEEEeqnarray}
for some $\tilde B>0$ and where $Q (\cdot)$ is the Q-function. To evaluate  $\Pr [\tilde i^{(\e)} \left ( \bar{ \vectsf S}_{\e,1},  \bar{ \vectsf S}_{\e,2};  \vectsf Y_c| \mathcal B_{\text{detect}} = B_{\text{dt}} \right) > \delta_{\e} ]$ we follow \cite[Lemma 47]{Yuri2012} which results in 
\begin{IEEEeqnarray}{rCl}
\lefteqn{\Pr [\tilde i^{(\e)} \left ( \bar{ \vectsf S}_{\e,1},  \bar{ \vectsf S}_{\e,2};  \vectsf Y_c| \mathcal B_{\text{detect}} = B_{\text{dt}} \right) > \delta_{\e}]} \notag \\
& \le& \frac{2}{M_{\e}M_{\s} n^{K_{\e} }} \left ( \frac{\log 2}{\sqrt{2\pi n}} + \frac{2 \tilde B} {\sqrt{n}}\right). \label{eq:83}
\end{IEEEeqnarray}

By combining \eqref{eq:71}, \eqref{eq:74}, \eqref{eq:79}, \eqref{eq:80}, \eqref{eq:82} and \eqref{eq:83} we have the following bound on the eMBB decoding error probability:
\begin{IEEEeqnarray}{rCl}
\lefteqn{\epsilon_{\e}^{(n)} \le \sum_{\Bdt}  \P_{\text{det}}^{|\Bdt|} (1-\P_{\text{det}})^{\eta-|\Bdt|} }\notag \\
&&\hspace{1.2cm}\Bigg [ \Delta_{\e}+ Q\left (\frac{-\log(M_{\e}M_{\s}) + n \tilde {\C}_{\e} - K_{\e} \log (n)}{\sqrt{n \tilde {\mathsf V}_{\e}}} \right) \Bigg], \IEEEeqnarraynumspace
\end{IEEEeqnarray}

where $\Delta_{\e}$ is defined in \eqref{eq:Deltae}.
According to Proposition~\ref{prop1}, it is required that $\epsilon_{\e}^{(n)}$ to be bounded above by $ \epsilon_{\e}$, i.e., $\epsilon_{\e}^{(n)} \le \epsilon_{\e}$. Thus
 \begin{IEEEeqnarray}{rCl}
\lefteqn{\epsilon_{\e} \ge \sum_{\Bdt}  \P_{\text{det}}^{|\Bdt|} (1-\P_{\text{det}})^{\eta-|\Bdt|} }\notag \\
&&\hspace{1.2cm}\Bigg [ \Delta_{\e}+ Q\left (\frac{-\log(M_{\e}M_{\s}) + n \tilde {\C}_{\e} - K_{\e} \log (n)}{\sqrt{n \tilde {\mathsf V}_{\e}}} \right) \Bigg]. \IEEEeqnarraynumspace
\end{IEEEeqnarray}
By taking the inverse of the $Q$-function, we have 
\begin{IEEEeqnarray}{rCl}
\lefteqn{\log M_{\e}}\notag \\
& \le& n \C_{\e} - \sqrt{n  \mathsf V_{\e}}Q^{-1}(\epsilon_{\e} - \Delta_{\e}) - K_{\e} \log (n) -\log M_{\s}, \IEEEeqnarraynumspace \label{eq:96}
\end{IEEEeqnarray}
where $\C_{\e}$ and $\mathsf V_{\e}$ are defined in \eqref{eq:Ce} and \eqref{eq:Ve}, respectively. Dividing both sides of \eqref{eq:96} by $n$ concludes the proof.

\end{document}